\documentclass[11pt]{article}
\usepackage{moriond,epsfig}

\def\be{\begin{equation}}
\def\ee{\end{equation}}
\def\bea{\begin{eqnarray}}
\def\eea{\end{eqnarray}}

\def \ptjet    {\ensuremath{ p_{T}^{\mathrm{jet}} }}
\def \ptbjet    {\ensuremath{ p_{T}^{b\mathrm{-jet}} }}
\def \ptz    {\ensuremath{ p_{T}^{Z} }}
\def \pt    {\ensuremath{ p_{T}}}
\def \vjets {$V+$\hspace{.5mm}jets}

\def \twoton {\ensuremath{ 2 \rightarrow N}}
\newcommand{\dzero}{D\O}
\newcommand{\epem}{\ensuremath{e^+ e^-}}

\newcommand{\mcfm}{{\sc mcfm}}
\newcommand{\pythia}{{\sc pythia}}
\newcommand{\herwig}{{\sc herwig}}

\newcommand{\sherpa}{{\sc sherpa}}
\newcommand{\alpgen}{{\sc alpgen}}

\begin{document}
\title{$W/Z+$\hspace{.5mm}Jets and $W/Z+$\hspace{.5mm}Heavy Flavor
  Jets at the Tevatron}

\author{Henrik Nilsen \footnote{For the \dzero\ and CDF 
    Collaborations.}}

\address{Albert-Ludwig University of Freiburg, Hermann-Herder-Str. 3,
  D-79104 Freiburg}

\maketitle

\vspace{-.5cm}
\begin{figure}[h!]
\begin{center}
\psfig{figure=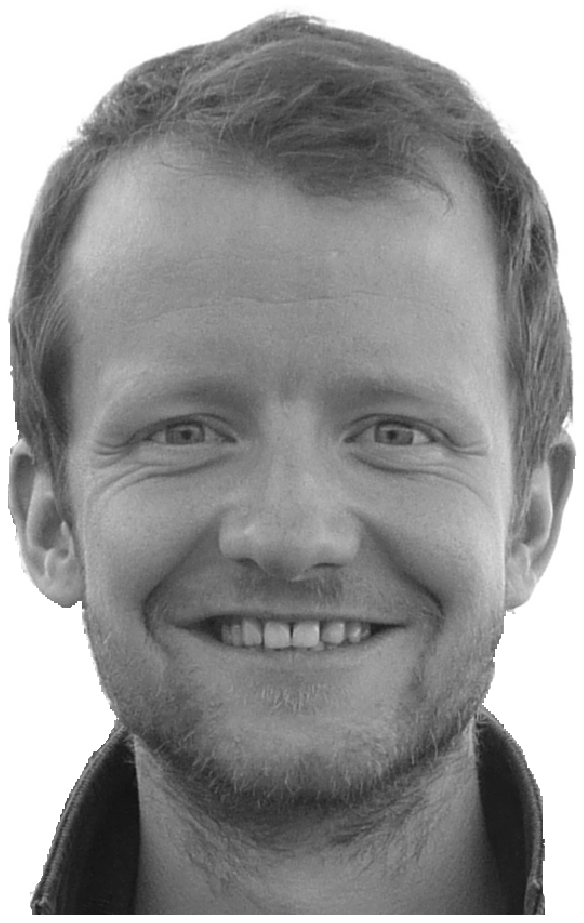,height=4cm}
\end{center}
\end{figure}
\vspace{-.2cm}

\abstracts{The associated production of jets and vector
  bosons is an important process at hadron colliders. An overview over
  recent Tevatron vector boson$+$jets measurements is given with an
  emphasis on comparisons between data and the predictions of various
  theory models.}

\section{Motivation}

The associated production of jets and vector bosons (\vjets) in hadron
collisions represents an important test of QCD. In addition, \vjets\ is
a significant source of background events in many measurements and
searches both at the Tevatron and the LHC. The development of
simulation codes which produce accurate predictions for \vjets\
production has been a very active field of research over the last few
years. The developments have followed two main paths: parton-level
fixed-order predictions with NLO accuracy;
and particle-level predictions from combining tree-level \twoton\
matrix elements with a parton shower algorithm.
These new models require validation against experimental measurements
of the properties of \vjets\ production. The leptonic decay modes
offer distinct experimental signals with low backgrounds, and during
the last two years a long list of \vjets\ measurements from the CDF
and \dzero\ experiments have been made public. All the measurements
presented here are fully corrected for detector effects, thus offering
a reference against which existing and future simulation models can be
validated and tuned. The measurements can be divided into those which
tag heavy-flavour (HF) jets and those which are inclusive in jet
flavour.

\section{$Z+$\hspace{.5mm}jets measurements}

\begin{figure}
\begin{center}
\psfig{figure=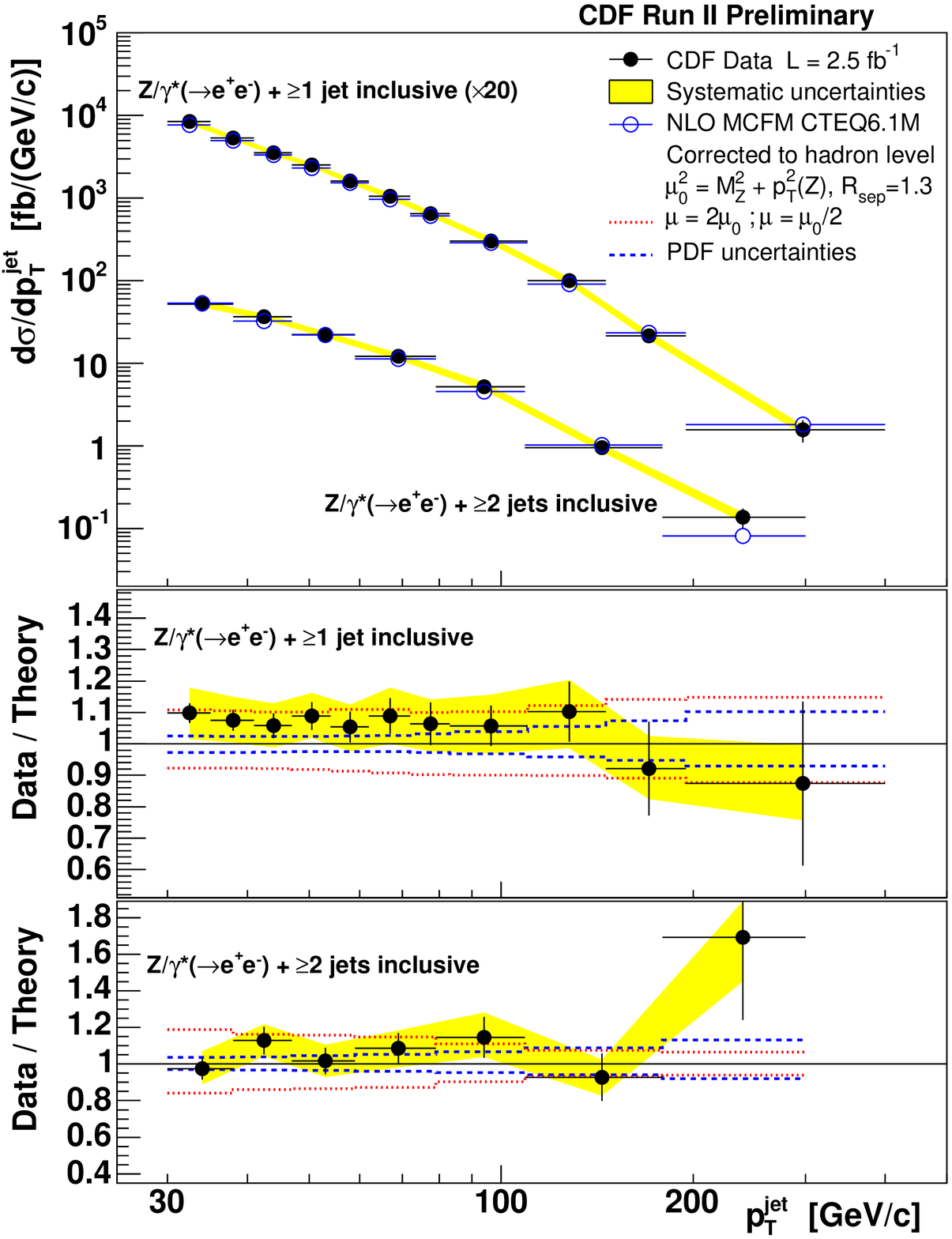,height=10cm}
\psfig{figure=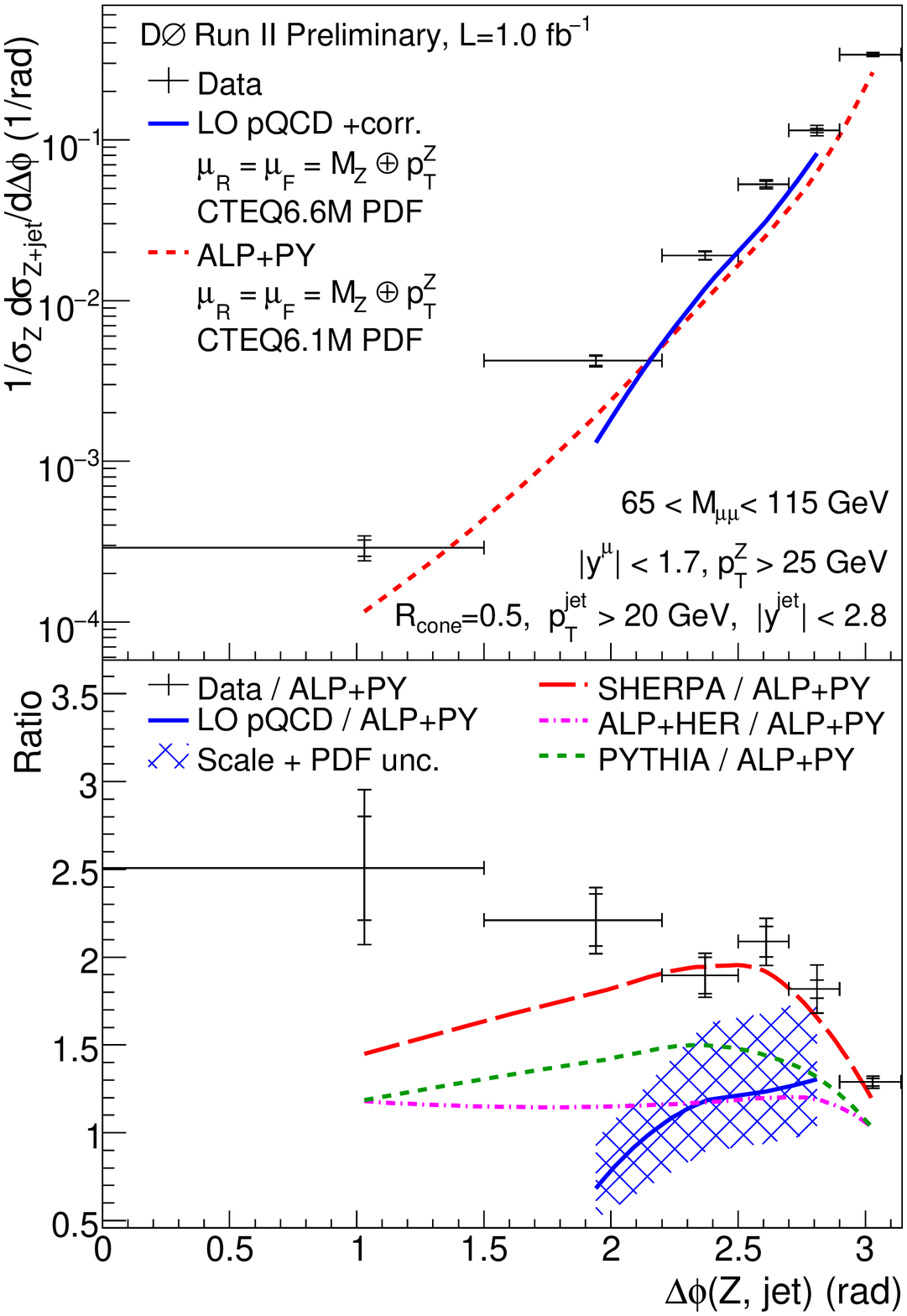,height=9.7cm}
\end{center}
\caption{Inclusive \ptjet\ spectra in $Z+N$-jet events, $N=1,2$, with
  data compared to NLO pQCD (right). Data compared with NLO pQCD and
  various event generators predictions for $\Delta\phi(Z,$jet$)$ in
  $Z+1$-jet events (left).}
\label{fig:cdf_zjets}
\end{figure}

CDF has presented measurements of the jet multiplicity in
$Z+$\hspace{.5mm}jets as well as the inclusive, differential \ptjet\
spectra in event with at least $N=1,2$ jets~\cite{cdfZjets}. The boson
is selected via its decay into an pair of high-$E_T$ electrons whose
invariant mass is compatible with $M_Z$. Jets are defined using the
Run~II mid-point algorithm and are required to satisfy $\pt > 30$~GeV
and $|y|<2.1$. The correction for detector effects is deduced from a
simulated event sample passed through a simulation of the detector. In
Fig.~\ref{fig:cdf_zjets} (left) the measured \ptjet\ spectra are
compared with parton-level NLO pQCD predictions from \mcfm~\cite{mcfm}
which have been corrected for hadronization and the underlying
event. The NLO predictions are seen to agree with data within
experimental and systematic uncertainties over one order of magnitude
in \ptjet\ and four orders of magnitude in cross section.

\begin{figure}
\begin{center}
\psfig{figure=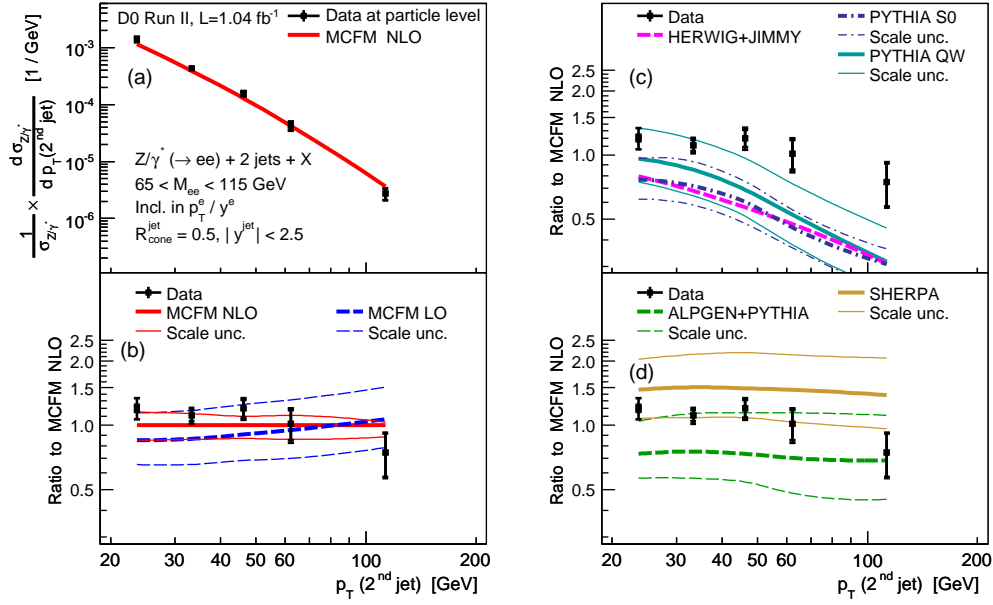,height=8cm}
\end{center}
\caption{Data compared with NLO pQCD and various event generators
  predictions for \pt($2^{nd}$ jet) in $Z+2$-jet events.}
\label{fig:d0_zjets}
\end{figure}

\dzero\ has presented measurements of the \ptjet\ spectra of the three
leading jets in the $Z(\rightarrow\epem)+$\hspace{.5mm}jets channel,
normalized to the inclusive $Z(\rightarrow\epem)$ cross
section~\cite{d0ZeeJets}. The event selection is similar to the CDF
analysis, with jets being reconstructed down to $20$~GeV. The
measurements are compared with both with fixed-order pQCD parton-level
predictions from \mcfm\ and the particle-level predictions of various
commonly used event generators. The comparisons for the second jet are
given in Fig.~\ref{fig:d0_zjets}. Both the LO and NLO pQCD predictions
are consistent with data within experimental and theoretical
uncertainties. As expected, the NLO prediction has significantly lower
scale uncertainties than the LO prediction, corresponding to a higher
predictive power. \pythia~\cite{pythia} using Tune A (``old''
$Q^2$-order parton shower) predicts less jet activity than seen in
data, and the discrepancies increase with \ptjet\ and jet
multiplicity. The same tendency is seen for
\herwig~\cite{herwig}. \pythia\ using Tune S0 (``new'' \pt-ordered
parton shower) gives good agreement for the leading \ptjet\ spectrum,
but no improvement over the old model for sub-leading jets. In
contrast, both \sherpa~\cite{sherpa} and \alpgen+\pythia~\cite{alpgen}
are found to predict the shapes of the \ptjet\ spectra reasonably well
for all three leading jets, with the latter generator giving somewhat
better agreement for the leading jet. The normalizations are affected
by significant scale uncertainties which increase with jet
multiplicity. \sherpa\ (\alpgen+\pythia) predicts more (less) jets
than observed in data, but for both codes the normalizations can be
made to agree with data by adjusting the choices of factorization and
renormalization scales.

Two \dzero\ studies~\cite{d0ZmumuJetI,d0ZmumuJetII} presents
measurements of the \pt\ and rapidity of the $Z$ and the leading jet,
as well as various angular correlations between the two objects. The
data are compared with NLO pQCD from \mcfm, \pythia\ using Tune A,
\sherpa, \alpgen+\pythia\ using Tune A, and, for the angular
correlation observables, \alpgen+\herwig. While fixed-order NLO
calculations are found found give accurate predictions for \pt\ and
jet multiplicity observables (see above), it does not describe the
spectrum of $\Delta\phi(Z,$jet$)$ (Fig.~\ref{fig:cdf_zjets} (right))
for values close to $\pi$, where multiple soft emissions are
important, or below $\sim 2$, where the underlying event gives sizable
contributions. Of the particle-level event generators, \sherpa\ is
found to give the most accurate description of the angular
correlations.

\section{$V+$\hspace{.5mm}HF-jet measurements}

Many searches for new particles, e.g.\ low-mass Higgs searches at the
Tevatron, tag $b$-jets in order to enhance the signal to background
ratio. In such searches, accurate predictions for the associated
production of a vector boson and heavy-flavour jets is of major
importance for the sensitivity of the analysis to new physics.

Both CDF and \dzero\ have presented measurements of a $W$ boson in
association with a single $c$ quark using similar
strategies~\cite{cdfWc,d0Wc}. This channel is sensitive to the
$s$-quark content of the proton at large $Q^2$, and it is a background
to top-quark measurements and searches for a low-mass Higgs particle
at the Tevatron. The $W$ is selected via a high-\pt\ lepton ($e$ or
$\mu$), and large missing $E_T$. A soft muon from a semi-leptonic
$c$-quark decay is used to tag $c$-jets. For signal events the two
leptons tend to have opposite charge, whereas the backgrounds show no
such charge correlation. CDF measures $(\sigma \times {\rm BR}) = 9.8
\pm 2.8$(stat)$\pm^{+1.4}_{-1.6}$(sys) pb, which is in good agreement
with the NLO pQCD prediction of $11 \pm^{+1.4}_{-3.0}$ pb. \dzero\
presents the differential \ptjet\ cross section for $W+c$ relative to
$W+$jet and sees agreement with \alpgen+\pythia\ within uncertainties.

Based on a similar event selection, CDF measures the $W+b$-jet cross
section~\cite{cdfWb}. The $W$ is selected via its decay into $e\nu$ or
$\mu\nu$, and a secondary-vertex algorithm is used to define a
$b$-quark enhanced sample. The $b$-quark content is extracted from the
secondary-vertex mass distribution by fitting with mass templates for
light-flavour, $c$ and $b$ quark samples. The cross section for
$\ptbjet > 20$~GeV is measured to be $(\sigma \times {\rm BR}) = 2.78
\pm 0.27$(stat)$\pm 0.42$(sys) pb. The \alpgen\ prediction of the
cross section is $0.78$ pb, which is a factor of $3-4$ below data, and
work is ongoing to understand this discrepancy.

A very similar $b$-tagging and $b$-content extraction technique is
used by CDF in an analysis~\cite{cdfZb} of $Z+b$-jet events in the
$ee$ and $\mu\mu$ channels. Cross sections are measured relative to
the inclusive $Z$ cross section and are presented differential in
$E_T^{b{\rm -jet}}$, $\eta^{b{\rm -jet}}$, \ptz\ and jet multiplicity
both for $b$ jets and flavour-inclusive jets. The total relative cross
section is measured to be $\sigma(Z+$jet$)/\sigma(Z) = (3.32 \pm
0.53$(stat)$\pm 0.42$(sys)$) \times 10^{-3}$. The NLO pQCD prediction
is $2.3 \times 10^{-3}$ for $\mu_F^2 = \mu_R^2 = m_Z^2 + p_{T,Z}^2$
and $2.8 \times 10^{-3}$ for $\mu_F^2 = \mu_R^2 =
\langle\ptjet\rangle^2$, in good agreement with data within
uncertainties. The prediction of \alpgen\ is $2.1 \times 10^{-3}$ and
\pythia\ predicts $3.5 \times 10^{-3}$. The large difference between
\alpgen\ and \pythia\ has been traced back to the higher choice of
scales used for \alpgen\ than for \pythia.
\begin{figure}
\begin{center}
\psfig{figure=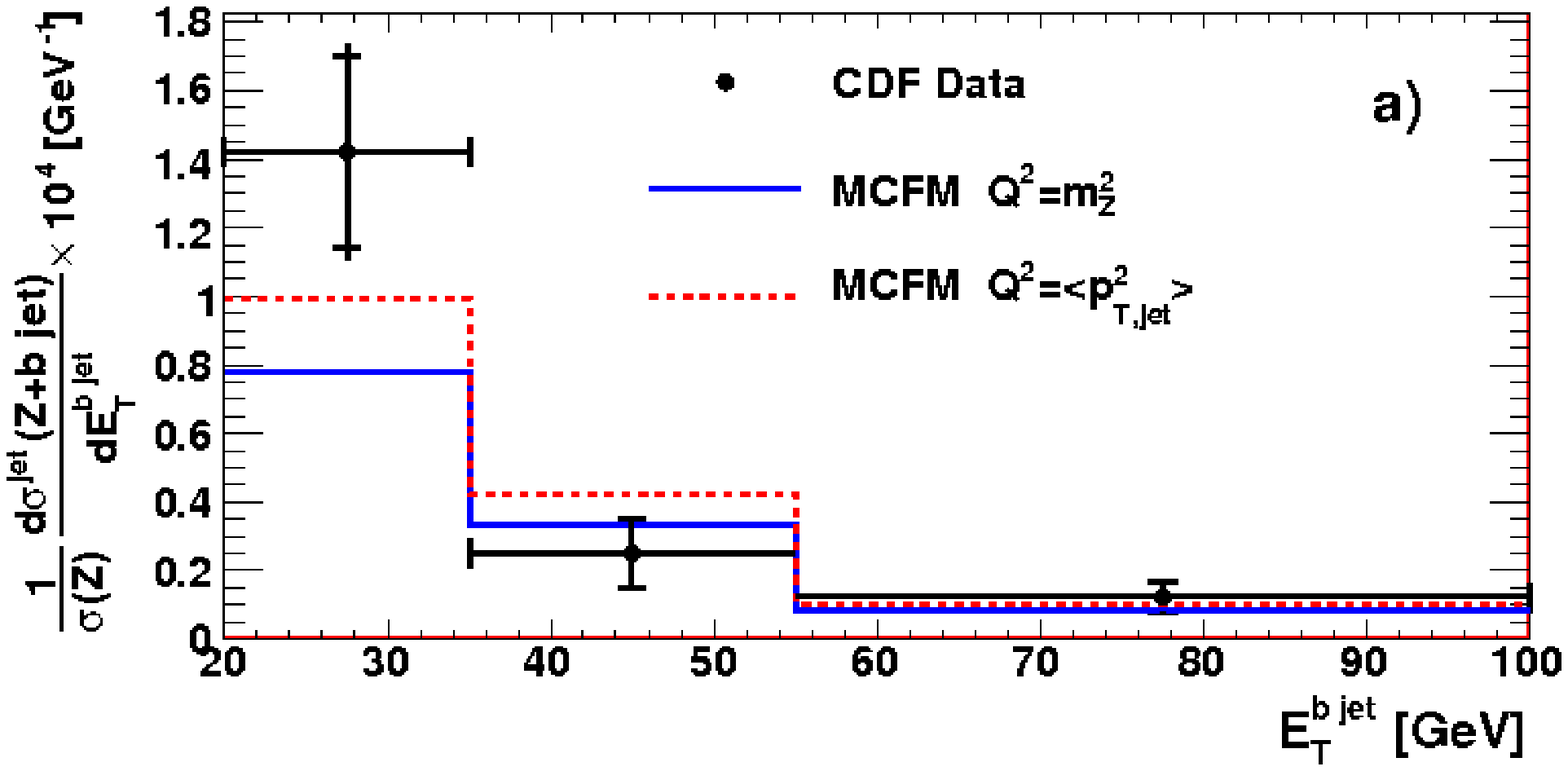,height=5cm,width=7.5cm}
\psfig{figure=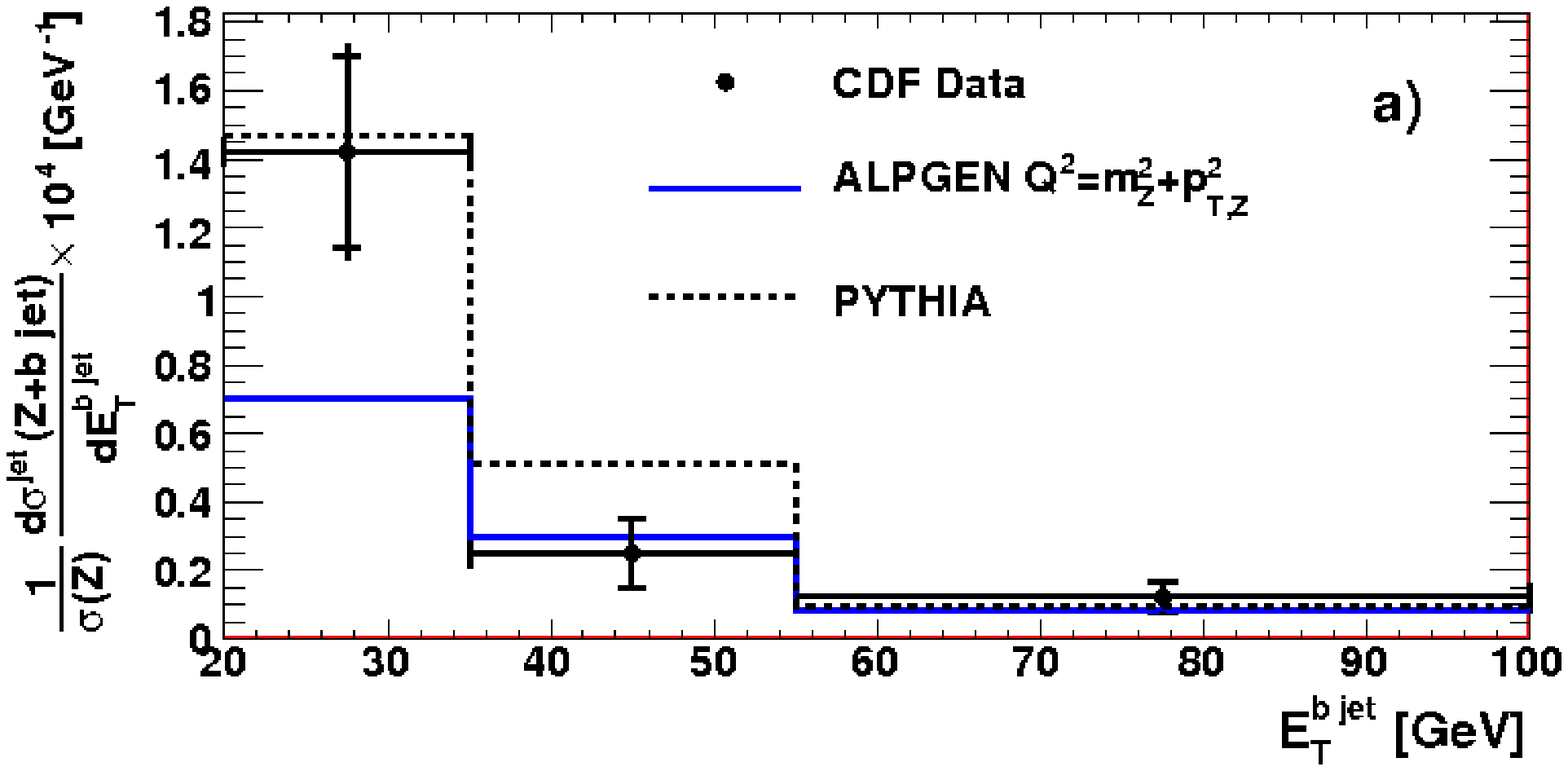,height=5cm,width=7.5cm}
\caption{The \ptbjet\ spectrum measured in $Z+b$-jet production
  compared with \mcfm, \pythia\ and \alpgen.}
\label{fig:cdf_zbjet}
\end{center}
\end{figure}

\vspace{-.1cm}

\section{Conclusions}

In addition to offering an important test of QCD, \vjets\ production
is a major source of background to many measurements and searches at
hadron colliders. Several new codes for simulating the associated
production of $Z/W$ and jets have become available over the last few
years, and the validation and tuning of these tools are of great
importance. A long list of \vjets\ measurements have become available
from the CDF and \dzero\ experiments during the last two
years. Parton-level predictions from NLO pQCD are found to offer the
highest predictive power for \ptjet\ spectra, showing good agreement
with data, both for flavor-inclusive and HF measurements. Generators
matching tree-level matrix elements with parton showers are found to
offer the most accurate particle-level predictions but have
significant scale uncertainties. Angular correlations show sensitivity
to multiple soft emissions and the underlying event and are therefore
partially outside of the scope of fixed-order pQCD calculations, and
event-generator predictions show varying agreement with data. In the
heavy-flavor channels, both pQCD and event-generator predictions are
found to be in agreement with data within uncertainties, with a
possible exception being $W+b$-jet production. Since all presented
measurements are fully corrected for detector effects they can be
directly used for testing and improving existing and future theory
models.


\vspace{-.2cm}
\section*{References}
\begin{thebibliography}{99}

\bibitem{cdfZjets} CDF Collaboration, T.~Aaltonen {\it et al.}, Phys.\
  Rev.\ Lett.\ {\bf 100}, 102001 (2008).

\bibitem{d0ZeeJets} \dzero~Collaboration, V.~M.~Abazov {\it et al.},
  submitted to Phys.\ Lett.\ B (2009).

\bibitem{d0ZmumuJetI} \dzero~Collaboration, V.~M.~Abazov {\it et al.}, Phys.\
  Lett.\ B {\bf 669}, 278 (2008).

\bibitem{d0ZmumuJetII} \dzero~Collaboration, V.~M.~Abazov {\it et
    al.}, \dzero\ Conference Note 5903-CONF.

\bibitem{cdfWc} CDF Collaboration, T.~Aaltonen {\it et al.}, Phys.\
  Rev.\ Lett.\ {\bf 100} 091803 (2008).

\bibitem{d0Wc} \dzero~Collaboration, V.~M.~Abazov {\it et al.}, Phys.\
  Lett.\ B {\bf 666} 23 (2008).

\bibitem{cdfWb} CDF Collaboration, T.~Aaltonen {\it et al.}, CDF
  Public Note 9321.

\bibitem{cdfZb} CDF Collaboration, T.~Aaltonen {\it et al.}, submitted
  to Phys.\ Ref.\ D (2008).

\bibitem{mcfm} J.~Campbell and R.~K.~Ellis, Phys.\ Rev.\ 
  D {\bf 65}, 113007 (2002).

\bibitem{pythia} T.~Sj\"{o}strand {\it et al.}, JHEP {\bf 0605}, 026
  (2006).

\bibitem{herwig} G.~Corcella {\it et al.}, JHEP {\bf 0101}, 010
  (2001).

\bibitem{sherpa} T.~Gleisberg {\it et al.}, JHEP {\bf 0902}, 007 (2009).

\bibitem{alpgen} M.~L.~Mangano {\it et al.}, JHEP {\bf 0307}, 001 (2003).

\end{thebibliography}
\end{document}